# Effects of the local structure dependence of evaporation fields on field evaporation behavior

Lan Yao[1], Travis Withrow[2], Oscar D. Restrepo[2], Wolfgang Windl[2], Emmanuelle A. Marquis[1*]

[1] Department of Materials Science and Engineering, University of Michigan, Ann Arbor, MI 48109

[2] Department of Materials Science and Engineering, The Ohio State University, Columbus, OH 43210

*Corresponding author

Phone: (734) 764 8717

Email: emarq@umich.edu

*Accurate three dimensional reconstructions of atomic positions, and full quantification of the information contained in atom probe tomography data relies on understanding the physical processes taking place during field evaporation of atoms from needle-shaped specimens. However, the modeling framework for atom probe tomography has remained qualitative at best. Building on the continuum field models previously developed, we introduce a more physical approach with the selection of evaporation events based on density functional theory calculations. This new model reproduces key features observed experimentally in terms of sequence of evaporation, desorption maps, and depth resolution, and provides insights into the physical limit for spatial resolution.*

Atom probe tomography, an advanced microscopy technique providing three-dimensional chemical information with near atomic resolution, relies on the fundamental principle of field evaporation of atoms from a needle-shape specimen subjected to a high electric field [1,2]. The chemical identity of each ion is inferred from time of flight measurements between the specimen and a 2D digital-detector positioned a few decimeters away from the specimen. The evaporated portion of the specimen is reconstructed into a 3D volume. The atoms are positioned one-by-one using a simple projection law to back-calculate the original position on the specimen surface from the detector coordinates of the ion hit while the depth assignment is incrementally calculated based on the sequence of evaporation [3]. This reconstruction method can theoretically yield accurate positioning of the atoms when the specimen has a spherical end cap and evaporation takes place one atomic layer at a time. However, spatial accuracy depends on a number of factors, which include local chemistry, microstructure, sequence of evaporation, and exact end shape of the specimen. Deviations from the idealized behavior assumed in reconstruction algorithms, i.e. layer-by-layer evaporation and spherical end cap, can lead to significant reconstruction artifacts. Numerous experimental evidence points to inaccurate representation of specimen shapes and evaporation sequence, particularly for materials with complex microstructures [4,5].

Quantification of the positioning errors and uncertainties remain a major challenge, and modeling of field evaporation, ion trajectories, and specimen shape evolution during evaporation would be a possible avenue to understand the evaporation process in more detail, explore resolution limits, correct for reconstruction errors, and generate more accurate reconstructed data. Vurpillot developed continuum simulations of field evaporation from needle-shaped specimens qualitatively reproducing some of the experimentally observed reconstruction artifacts such as those observed during evaporation of small precipitates or multilayer structures [6,7]. Expanding on these initial calculations, Oberdorfer developed field evaporation simulation from realistic scale samples [8]. Unfortunately, these simulations do not produce some of the evaporation patterns observed experimentally. As an example, the predicted evaporation pattern formed by the cumulative hits on the detector obtained from a <001> crystal (e.g. Fig. 4 of Ref. [8]) is different from the one observed experimentally (e.g. Fig. 6 of Ref [9]), particularly in the density patterns around the (100) pole. Notably, these simulations select the evaporating atom based on the highest value of the applied field, not taking into account its explicit adhesion energy, which depends on its local atomic environment on the surface. As we will show in this letter, the criterion by which evaporating atoms are selected in the simulations is a crucial step, which not only determines the evaporation sequence, but also modifies the morphology of the specimen surface and therefore the ion trajectories and their hit position on the detector. We therefore propose here a more physical evaporation criterion, which explicitly considers local atomic arrangements by calculating not only the strength of the applied electrical field but also the variations of the adhesion energy and thus evaporation field with local structure.

The field and trajectory simulations are performed following the same principles as the previously published evaporation models [10,11]. The specimen is represented by a 16 nm diameter cylinder with a spherical end cap. Atoms are positioned on an fcc crystal lattice and the <001> direction is oriented parallel to the cylinder axis. The specimen is subjected to a potential of 1 kV. The actual field distribution at and near the specimen surface is solved using Poisson's equation under the assumption of constant surface charge on the sample tip. The evaporating atom is selected as discussed below and the electrical field is recomputed before continuing to the next evaporation event. Finally, the ion trajectories are computed to determine the impact of the ion on the detector. If a full trajectory computation within the electric field distribution were to be performed, it would reproduce the curved trajectories that are observed experimentally as previously shown [12]. However in this work, calculation speed was favored and the trajectories are simplified to straight lines following the direction of the electric field computed at the location of the evaporating atom on the surface, with a detector positioned at 10 nm away form the specimen. While the chosen simplification modifies the magnification distribution over the specimen surface, it does not affect our current purpose and conclusions focusing on the evaporation behavior.

To select the evaporation events, we start from the assumption of thermally activated evaporation where the atom with the shortest evaporation time $\tau$ is the one evaporated. Under the assumption of harmonic transition state theory, the evaporation time is given by

$$\tau = \tau_0 \exp(Q/k_B T), \qquad (1)$$

where the activation energy $Q$ depends on the binding energy $Q_0$ and through the Schottky term on the applied field $F$,

$$Q = Q_0 - \sqrt{(ne)^3 F}. \qquad (2)$$

The temperature, $T$, is commonly assumed to be low enough that the lowest-barrier atom is evaporated, which is the one with the smallest value for $Q$. The fact that the evaporation time $\tau$ in Eq. (1) for an ion with charge $n+$ is only finite for $T \rightarrow 0$ if $Q \leq 0$ defines a minimum value for $F$ called evaporation field $F_0^{n+}$, which, using Eq. (2), is given by

$$F_0^{n+} = (ne)^{-3} \left(Q_0^{n+}\right)^2. \qquad (3)$$

Inserting this into Eq. (2) and assuming finite, but small temperature, means the selection criterion for the fastest evaporating atom from all atoms is

$$\max\left(\sqrt{F} - \sqrt{F_0^{n+}}\right). \quad (4)$$

In the simplest and currently exclusively used approximation, which we call constant-binding energy or $Q_{\text{const}}$-model, $Q_0^{n+}$ (or equivalently $F_0^{n+}$) is assumed to be the same for all atoms of the same species, resulting in the selection of the atom with the highest local field $F$. To provide a more realistic description of the evaporation events, we introduce a local binding energy or $Q_{\text{loc}}$-model, where $Q_0^{n+}$ depends on the local environment of the evaporating atom, which we determine from density-functional theory (DFT) calculations. Within the so-called Müller equation [13], which is based on work by Smith [14] and further developed by Forbes [15],

$$Q_0^{n+} = H_b + I_{n+} - n\varphi, \quad (5)$$

where $H_b$, $I_{n+}$, and $\varphi$ are heat of evaporation, ionization energy, and local work function, respectively. Previous DFT work [16] has shown that the Müller model gives reliable evaporation field values, specifically for the case of elemental Al considered here, even though it miscalculates the escape surface positions.

We calculate $H_b$, $I_{n+}$, and $\varphi$ by DFT in VASP [17] using PAW potentials with GGA(PW91) exchange-correlation functionals [18]. All calculations use a cutoff energy of 240 eV and a 4×4×1 k-point mesh. Simulation cells are relaxed fcc 3×3×3 supercells for (100) and 3×3×2 and 4×4×2 supercells for (111), each with 15 Å vacuum. Heats of evaporation are calculated as the enthalpy difference between a cell with an adatom at half vacuum and the relaxed reference cell with the adatom at the surface. A large number of surface configurations and adatom environments are examined, sufficient to include all sensible nearest, second-nearest and third-nearest neighbor configurations for an atom on any surface. The local work function is calculated by subtracting the calculated Fermi energy from the electrostatic potential at half vacuum. Ionization energy is calculated by taking the difference between the energies of neutral and charged atoms in vacuum and then fitting $E(a) = E_\infty - E'/a$ to calculations for cells with side lengths between 15 and 30 Å. As expected, calculated work function and ionization energies agree well with experimental results with $\varphi_{calc} = 4.23$ eV, $I_{al+, calc} = 6.04$ eV and $I_{al2+, calc} = 24.87$ eV compared to 4.24 eV, 5.99 eV and 24.81 eV respectively [19]. Calculated values for the evaporation fields are 1.5 – 1.7 V/Å for an isolated Al adatom evaporating with a single charge and 1.80 – 2.11 V/Å for $Al^+$ at step edges. The calculated values for surface features on [100] and [111] Al-surfaces are shown in **Fig. 1(a)** and **(b)**. Averaging over the evaporation fields of evaporated atoms in simulation as described below results in values of 2.0 eV for both $Q_{const}$ and $Q_{loc}$ models. Since the calculated evaporation fields for $Al^{2+}$ are 3.3-3.5 V/Å, only single charge ions should evaporate in the case of $T \rightarrow 0$, which is therefore the only charge state considered in the following. The calculated values are in good agreement with the traditionally used single values of 1.9 V/Å for $Al^+$ and 3.5 V/Å for $Al^{2+}$ [20] and experimentally determined values between 1.5 and 3.0 V/Å [21].

The $Q_{loc}$-model, based on tabulated evaporation-field values for all sensible environments up to third neighbors for surface atoms, was compared to the $Q_{const}$-simulations in terms of the evaporation behaviors of crystallographic facets, the resulting field desorption maps formed by the cumulating ion positions on the detector, and the reconstructed volumes created using a traditional reconstruction algorithm.

The simulated evaporating specimen surfaces determined with both models can be visualized in Movies 1 and 2 (Multimedia view) and representative still images are shown in

**Fig. 2**. The surface atoms are color-coded according to the value of F for the $Q_{const}$-model and $\left(\sqrt{F} - \sqrt{F_0^{n+}}\right)$ for the $Q_{loc}$- model. It is not surprising that both models highlight the atomic terraces produced on the evaporated surfaces, as these locations tend to correspond to the highest applied field and the least binding sites. A clear difference is the shape of the low index planes during evaporation, particularly the {111} planes in the $Q_{loc}$- model that exhibit a 6-fold symmetry facetted morphology was observed experimentally [22] but is not visible in the $Q_{const}$- model. This can be easily understood from the evaporation fields at the edges of surface islands shown in **Fig. 1(a)** and **(b)**, which vary strongly with edge orientation (correlated to the number of neighbors that an edge atom has) and thus result in high-evaporation and low-evaporation directions.

Another way to probe the difference in surface morphologies is to examine the distribution of the evaporation field strengths based on the surface configurations that developed during evaporation. **Figure 1(c)** shows the distributions calculated for 92,000 consecutively evaporating ions. Of course, the evaporation field in the $Q_{const}$-model is kept constant during evaporation, therefore this comparison only highlights the morphological differences between the evaporating surfaces. The distribution generated from the $Q_{loc}$-model shows that the majority of evaporation events shift to edges where edge atoms have lower numbers of neighbors and thus lower evaporation fields (e.g. the atoms on the [110] edge of the surface island in **Fig. 1(a)**). Lower-neighbor edge evaporation nearly doubles, while higher-neighbor edge evaporation decreases by 40%. The $Q_{loc}$ sequence also produces only half as many corner or isolated-adatom evaporation events and completely eliminates very-high-evaporation-field events. As described above, a prominent result of the evaporation sequence changed in this way is faceting of evaporating surface islands.

Because of these differences in specimen surface morphology and sequence of evaporation, the resulting field distributions near the specimen surface differ between the two models. These differences in the field distribution modify the trajectories of the evaporated ions and result in different ion detector positions. The cumulative ion positions on the detector form so-called desorption maps that are calculated for the two different models and shown in **Figure 3**. Not surprisingly, the $Q_{const}$-desorption map (**Fig. 3a**) reproduces the features previously published using the same evaporation criterion. Those are the low density regions at the center of the [001] and [111] poles, and the 4-fold symmetry pattern formed by low density zone lines connecting the [001] and [111] poles. These zone lines also represent the intercept between the specimen surface and the {110} planes. Within the low-density lines, one can also see two parallel rows of atoms. In contrast, a richer pattern develops in the $Q_{loc}$- desorption map (**Fig. 3b**). The [001] pole has a "flower"-like pattern with higher density petals along the {011} zone lines. A similar orientated "flower" pattern can be observed in experimental maps obtained by field desorption of Al [9]. However, differences with the experimental data are also noticed. In particular, the simulated density patterns do not reproduce the experimentally observed low-density regions at low index poles and along zone lines. First, the model considers smaller specimens than used for experimental samples, which leads to a loss of details on the higher order poles. This will be remediated when using real scale simulations. Second, the simplification regarding the trajectory calculations (omission of the curvature of the ion trajectories and consequently of the image compression [23], and the significantly shorter flight path of 20 nm, compared to a 100 mm experimental flight path length) results in a decrease of the locally enhanced magnification that low index surfaces normally exhibit. Therefore, the low-density at zone-line regions can hardly be observed in the current simulation. Again, this can be addressed by doing full trajectory calculations. Finally and probably most importantly, the possible movement of atoms on the surface of the specimen is not currently considered in our deterministic approach, such as the so-called roll up motion where kink atoms move over a ledge as they evaporate [24], surface rearrangement of kink atoms [25], or adatom surface diffusion [26] would also account for the local density variations that also exhibit a crystallographic dependence.

Beyond difference in the desorption maps on the detector, another consequence of the differences in the evaporation sequence between the two models can be observed once the positions of the ions are reconstructed into a 3D volume. We note that the reconstruction algorithm itself may introduce additional artifacts, however it provides a useful visualization of depth resolution. Reconstruction of the simulated data follows a standard reconstruction protocol used for atom probe tomography data [27]. The lateral positions are calculated from the coordinates of the ion hits on the detectors assuming a simple projection law. The depth assignment is linked to the sequence of evaporation in that after every evaporation event, depth is incremented by a small amount determined so that the overall atomic density of the reconstructed volume agrees with the density of the material being reconstructed. The depth increment is calculated from the projection law, the atomic density of the analyzed material, and the position of each ion on the specimen surface. **Figure 4** displays slices taken from data reconstructed in this way that was generated using both models. The slices contain the [002] and [111] poles and the lattice planes are clearly resolved in the different orientations. The $Q_{loc}$-model yields significant blurring of the lattice planes compared to the $Q_{const}$-model. The blurring or loss of spatial resolution results from the evaporation events being less localized and correlated and more spread over the specimen surface due to the local variation of $F_0^{n+}$ and the algorithm by which a depth coordinate is assigned to each evaporated ion. In the $Q_{\text{const}}$-model, atomic terraces evaporate readily where the external field is strongest with limited evaporation events taking place elsewhere on the surface. The atoms pertaining to the same atomic terrace are therefore reconstructed with close depth coordinates. In the $Q_{\text{loc}}$-model, while atomic terraces evaporate, evaporation events also take place elsewhere on the specimen surface where $F_0^{n+}$ is lower, leading to a wider range of depth values being assigned to the atoms pertaining to the same atomic plane or terrace. A similar blurring of atomic planes is observed in the experimental data, specifically from Al in a similar [111] orientation, e.g. Ref. [9].

In summary, we have shown that by explicitly evaluating the evaporation field as a function of local structure at each atomic site, the evaporation model is improved and a number of previously unaccounted field desorption and APT "artifacts" can be understood. Using local evaporation fields leads to a more random evaporation process than previously modeled where a constant evaporation field strength was used for all atoms. It also suggests an inherent limitation to spatial resolution in the context of the current reconstruction algorithms. The proposed model provides an opportunity to simulate the evaporation behavior of more complex materials and microstructures, such as alloys where each element has an evaporation field value that is not only dependent on surface structure but also dependent on the chemistry of its neighbors. Future steps will address the evaporation and spatial accuracy of the atomic reconstructions from structural defects, e.g. grain boundaries, dislocations, and precipitates, where the disruption of the atomic structure and inhomogeneous chemistry have been used to qualitatively explain the distinct experimental evaporation patterns.

**Acknowledgment**

The authors wish to acknowledge financial support from the Air Force Office of Scientific Research Award No. FA9550-14-1-0249 and computational support by the Ohio Supercomputer Center under Grant No. PAS0072.

**Figure and movie captions**

**FIG. 1** Calculated evaporation fields for different surface features for (a) [100] and (b) [111] surfaces of Al. (c) Distribution of the calculated evaporation field strengths for 92,000 consecutive evaporation events using the surface configurations developing during $Q_{\text{const}}$-model (closed red points) or the $Q_{\text{loc}}$-model (open black circles).

**FIG. 2** Simulated evaporated surfaces using (a) the $Q_{\text{const}}$- model and (b) the $Q_{\text{loc}}$- model.

**FIG. 3** Desorption maps simulated using (a) the $Q_{\text{const}}$- model and (b) the $Q_{\text{loc}}$- model.

**FIG. 4** Depth slices taken from the simulated data after 3D reconstruction of the first 92,000 consecutive evaporation events and containing <001> and <111> poles. (a) $Q_{\text{const}}$-model (b) $Q_{\text{loc}}$-model

**Movie 1**: Evaporation sequence of 92,000 atoms from the surface of a specimen using the $Q_{\text{const}}$- model.

**Movie 2**: Evaporation sequence of 92,000 atoms from the surface of a specimen using the $Q_{loc}$-model.

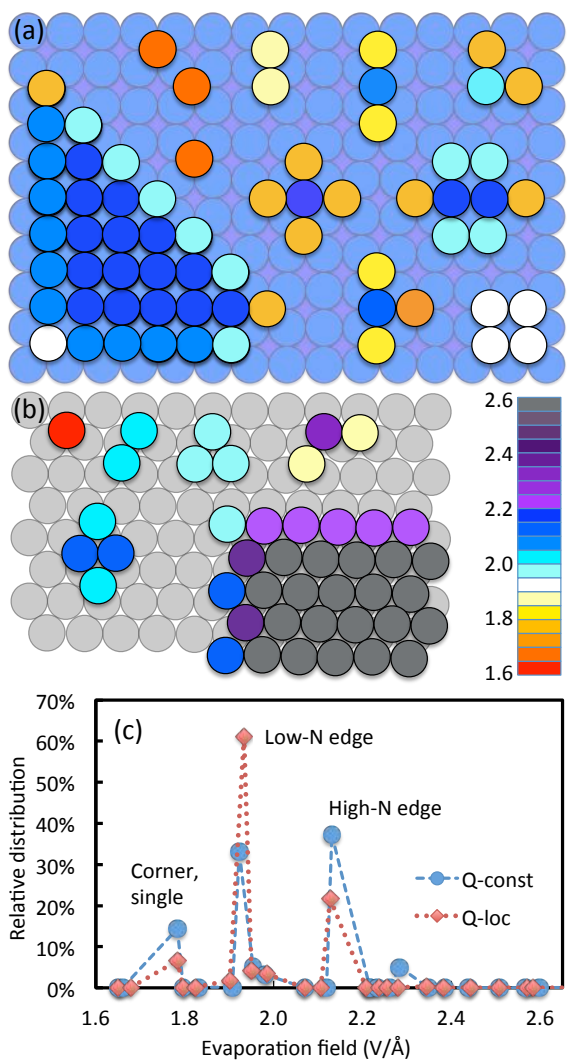

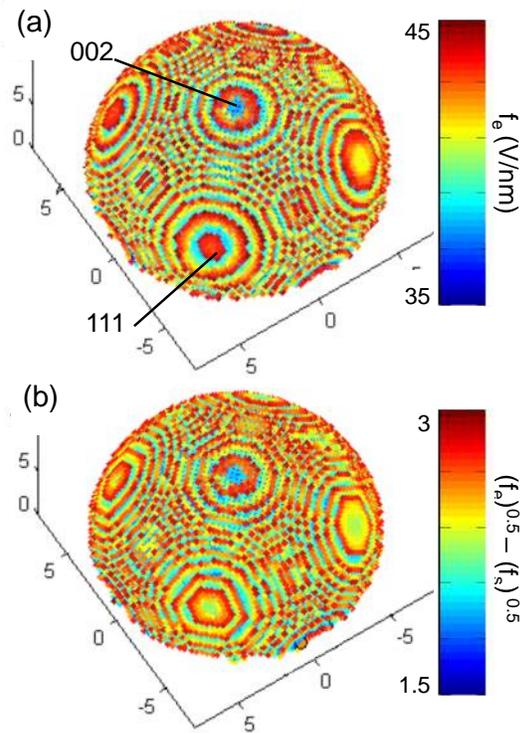

FIG. 2

FIG.1

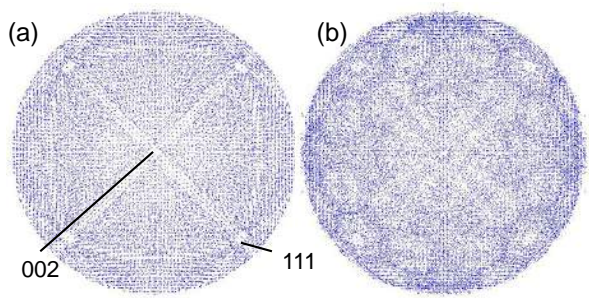

FIG. 3

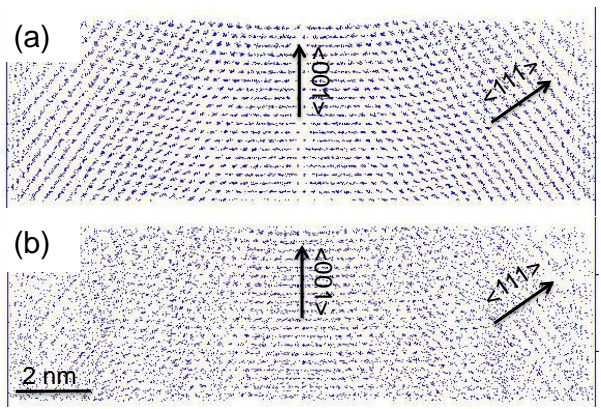

FIG. 4